\documentstyle[12pt]{article}

\def\noi{\noindent}

\def\nqq{\hspace{-2em}}

\def\barr{\left(\begin{array}}
\def\earr{\end{array}\right)}
\def\beq#1{\begin{equation}\label{#1}}
\def\eeq{\end{equation}}
\def\ber#1{\begin{eqnarray}\label{#1} \nqq}
\def\eer{\end{eqnarray}}
\def\eern{\nonumber \end{eqnarray}}
\def\nn{\nonumber}
\def\mm{\\ \nqq}
\newcommand{\bear}[1]{\begin{eqnarray}\label{#1}}
\newcommand{\ear}{\end{eqnarray}}

\catcode`\@=11 \@addtoreset{equation}{section}\catcode`\@=12

\newcommand{\R}{\mbox{\bf R}}

\newcommand{\N}{\mbox{\bf N}}

\newcommand{\sign}{\mathop{\rm sign}\nolimits}

\newcommand{\sh}{\mathop{\rm sh}\nolimits}
\newcommand{\ch}{\mathop{\rm ch}\nolimits}

\newcommand{\eps}{\varepsilon}
\newcommand{\tri}{\triangle}

\newcommand{\p}{\partial}

\newcommand{\fnm}{\footnotemark}
\newcommand{\fnt}{\footnotetext}

\begin{document}

\begin{center}
\large\bf
BLACK HOLE  P-BRANE SOLUTIONS FOR GENERAL
INTERSECTION RULES
\\[15pt]
\normalsize\bf V.D. Ivashchuk\fnm[1]\fnt[1]{ivas@rgs.phys.msu.su},
and
V.N. Melnikov
\fnm[2]\fnt[2]{melnikov@rgs.phys.msu.su} \\[10pt]

\it Center for Gravitation and Fundamental Metrology,
VNIIMS, 3/1 M. Ulyanovoy Str.,
Moscow 117313, Russia  and\\
Institute of Gravitation and Cosmology, PFUR,
Michlukho-Maklaya Str. 6, \\ Moscow 117198, Russia

\end{center}

\vspace{15pt}

\small\noi

\begin{abstract}

Black hole  generalized $p$-brane  solutions  for
a wide class of intersection rules are obtained.
The solutions are defined on a manifold that contains a product
of $n - 1$ Ricci-flat ``internal'' spaces.
They are defined up to  a set of functions $H_s$ obeying a
non-linear differential equations
(equivalent to Toda-type equations) with
certain boundary conditions. A conjecture on polynomial
structure of governing functions $H_s$ for intersections related to
semisimple Lie algebras is suggested. This conjecture is proved for
Lie algebras: $A_m$, $C_{m+1}$, $m \geq 1$.
Explicit formulas for  $A_2$-solution are obtained.  Two examples of
$A_2$-dyon solutions (e.g. dyon in $D = 11$ supergravity and  Kaluza-Klein
dyon) are considered.  Post-Newtonian parameters $\beta$ and $\gamma$
corresponding to $4$-dimensional section of the metric are calculated.
It is shown that  $\beta$ does not depend
upon intersections of $p$-branes.
Extremal black hole configurations are also considered.

\end{abstract}

\vspace{10cm}

\pagebreak

\normalsize

\section{Introduction}

 At present there exists an interest to the so-called $M$-theory
(see, for example, \cite{M-th1}-\cite{M-th2}).
This theory is ``supermembrane'' analogue of
superstring models \cite{GSW} in $D=11$. The low-energy limit of
$M$-theory after a dimensional reduction leads to models governed by a
Lagrangian containing a metric, fields of  forms and scalar fields.
These models contain a large variety of the so-called
$p$-brane solutions (see \cite{St}-\cite{IK} and references
therein).

In \cite{IMC} it was shown that after
the dimensional reduction on the
manifold $M_0\times M_1\times\dots\times M_n$  when the composite
$p$-brane ansatz for fields of forms
is considered the problem is reduced to the gravitating
self-interacting $\sigma$-model with certain constraints imposed. (For
electric $p$-branes see also \cite{IM0,IM,IMR}.) This representation
may be considered as a tool for obtaining different solutions
with intersecting $p$-branes. In \cite{IMC,IMR,IMBl,IKM,GrI}
the Majumdar-Papapetrou type solutions (see \cite{MP})
were obtained (for non-composite
case see \cite{IM0,IM}). These solutions corresponding to Ricci-flat
factor-spaces $(M_i,g^i)$, ($g^i$ is metric on $M_i$)
$i=1,\dots,n$, were also
generalized to the case of Einstein internal spaces \cite{IMC}.  Earlier
some special classes of these solutions were considered in
\cite{Ts1,PT,GKT,AR,AEH,AIR}. The obtained solutions take place, when
certain (block-)orthogonality relations (on couplings parameters,
dimensions of "branes", total dimension) are imposed. In this situation a
class of cosmological and spherically-symmetric solutions was obtained
\cite{IMJ,Br2,IMJ2}. Special cases were also considered in
\cite{LPX,BGIM,GrIM,BKR}.  The solutions with the horizon were studied
in details in \cite{CT,AIV,Oh,IMJ,BIM,IMBl,Br2,CIM}.

In models under consideration  there exists a large variety of
Toda-chain solutions,  when certain intersection rules are satisfied
\cite{IMJ}.  Cosmological and spherically symmetric solutions with
$p$-branes and $n$ internal spaces  related to $A_m$ Toda chains
were previously  considered in   \cite{LPX,LMPX} and   \cite{GM1,GM2}.
Recently in \cite{IK} a family of $p$-brane  solutions
depending on  one harmonic function with next to arbitrary
(up to some restrictions) intersection rules  were obtained.
These solutions are defined up to
solutions of Laplace and Toda-type equations and correspond to
null-geodesics of the sigma-model target-space metric.

Here we consider a family of spherically-symmetric and
cosmological type solutions from \cite{IK}
(see Sect. 2) and single out a subclass of black-hole configurations
related to Toda-type equations with
certain asymptotical conditions  imposed (Sect. 3).
These black hole solutions are
governed by functions $H_s = H_s(R)$ obeying a set of second order
non-linear differential equations  with some boundary relations
imposed. For positively defined
scalar field metric $(h_{\alpha \beta})$ all
$p$-branes in this solution should contain a time manifold
(see Proposition 1 from Sect. 3).
This agrees with Theorem  3 from Ref. \cite{Br2}
(for orthogonal case, see also \cite{BIM}).

In Sect. 4 we suggest a hypothesis: the functions $H_s$
are polynomials when intersection rules correspond to
semisimple Lie algebras.  This hypothesis (Conjecture 1 )
It is also confirmed by special  black-hole ``block orthogonal''
solutions considered earlier in \cite{Br,IMBl,CIM,IMJ2}.
An analogue of this conjecture for extremal black holes
was considered earlier in \cite{LMMP} (see Conjecture 2 in Sect. 7).

In Sect. 5  explicit formulas for the solution
corresponding to the algebra $A_2$ are obtained.  These formulas are
illustrated by two examples of $A_2$-dyon solutions: a dyon in $D = 11$
supergravity and Kaluza-Klein dyon.
In Sect. 6 post-Newtonian   parameters  $\beta$ and $\gamma$ corresponding
to $4$-dimensional section of the metric are calculated (for
special ``block orthogonal'' case see \cite{IMJ2,CIM}).
In Sect. 7 extremal black hole solutions  are considered.
These solutions forms  a subclass
in a class of more general solutions
depending on one harmonic function obtained recently in
\cite{IK}.


\section{Solutions with $p$-branes depending on one variable}

We consider a  model governed by
the action \cite{IMC}
\ber{1.1}
S=\int d^Dx \sqrt{|g|}\biggl\{R[g]-h_{\alpha\beta}g^{MN}\p_M\varphi^\alpha
\p_N\varphi^\beta-\sum_{a\in\tri}\frac{\theta_a}{n_a!}
\exp[2\lambda_a(\varphi)](F^a)^2\biggr\}
\eer
where $g=g_{MN}(x)dx^M\otimes dx^N$ is a metric,
$\varphi=(\varphi^\alpha)\in\R^l$ is a vector of scalar fields,
$(h_{\alpha\beta})$ is a  constant symmetric
non-degenerate $l\times l$ matrix $(l\in \N)$,
$\theta_a=\pm1$,
\beq{1.2a}
F^a =    dA^a
=  \frac{1}{n_a!} F^a_{M_1 \ldots M_{n_a}}
dz^{M_1} \wedge \ldots \wedge dz^{M_{n_a}}
\eeq
is a $n_a$-form ($n_a\ge1$), $\lambda_a$ is a
1-form on $\R^l$: $\lambda_a(\varphi)=\lambda_{\alpha a}\varphi^\alpha$,
$a\in\tri$, $\alpha=1,\dots,l$.
In (\ref{1.1})
we denote $|g| =   |\det (g_{MN})|$,
\beq{1.3a}
(F^a)^2_g  =
F^a_{M_1 \ldots M_{n_a}} F^a_{N_1 \ldots N_{n_a}}
g^{M_1 N_1} \ldots g^{M_{n_a} N_{n_a}},
\eeq
$a \in \tri$. Here $\tri$ is some finite set.

Let us consider a family of one-variable sector
solutions to field equations corresponding to the action
(\ref{1.1}) and depending upon one variable $u$
\cite{IK}. These solutions are defined on the manifold
\beq{1.2}
M =    (u_{-}, u_{+})  \times
M_1  \times M_2 \times  \ldots \times M_{n},
\eeq
where $(u_{-}, u_{+})$  is  an interval belonging to $\R$.
The solutions read \cite{IK}
\bear{1.3}
g= \biggl(\prod_{s \in S} [f_s(u)]^{2 d(I_s) h_s/(D-2)} \biggr)
\biggl\{[f_1(u)]^{2d_1/(1-d_1)}\exp(2c^1u + 2 \bar c^1)\\ \nn
\times[w du \otimes du+ f_1^2(u)g^1] +
\sum_{i = 2}^{n} \Bigl(\prod_{s\in S}
[f_s(u)]^{- 2 h_s  \delta_{i I_s} } \Bigr)
\exp(2c^i u+ 2 \bar c^i) g^i\biggr\}, \\ \label{1.4}
\exp(\varphi^\alpha) =
\left( \prod_{s\in S} f_s^{h_s \chi_s \lambda_{a_s}^\alpha} \right)
\exp(c^\alpha u + \bar c^\alpha), \\ \label{1.5}
F^a= \sum_{s \in S} \delta^a_{a_s} {\cal F}^{s},
\ear
$\alpha=1,\dots,l$.
In  (\ref{1.3})  $w = \pm 1$,
$g^i=g_{m_i n_i}^i(y_i) dy_i^{m_i}\otimes dy_i^{n_i}$
is a Ricci-flat  metric on $M_{i}$, $i=  2,\ldots,n$,
the space  $(g^1, M_{1})$ is an Einstein space
of non-zero curvature:
\beq{1.5a}
R_{m n}[g^1 ] = \xi^1 g^1,
\eeq
$\xi^1 \neq 0$, and
\beq{1.11}
\delta_{iI}=  \sum_{j\in I} \delta_{ij}
\eeq
is the indicator of $i$ belonging
to $I$: $\delta_{iI}=  1$ for $i\in I$ and $\delta_{iI}=  0$ otherwise.

The  $p$-brane  set  $S$ is by definition
\ber{1.6}
S=  S_e \cup S_m, \quad
S_v=  \cup_{a\in\tri}\{a\}\times\{v\}\times\Omega_{a,v},
\eer
$v=  e,m$ and $\Omega_{a,e}, \Omega_{a,m} \subset \Omega$,
where $\Omega =   \Omega(n)$  is the set of all non-empty
subsets of $\{ 2, \ldots,n \}$. Hence
all $p$-branes do not ``live'' in  $M_1$.

Any $p$-brane index $s \in S$ has the form
\ber{1.7}
s =   (a_s,v_s, I_s),
\eer
where
$a_s \in \tri$, $v_s =  e,m$ and $I_s \in \Omega_{a_s,v_s}$.
The sets $S_e$ and $S_m$ define electric and magnetic $p$-branes
correspondingly. In
(\ref{1.4})
\ber{1.8}
\chi_s  =   +1, -1
\eer
for $s \in S_e, S_m$ respectively.
In (\ref{1.5})  forms
\beq{1.9}
{\cal F}^s= Q_s
\left( \prod_{s' \in S}  f_{s'}^{- A_{s s'}} \right) du \wedge\tau(I_s),
\eeq
$s\in S_e$, correspond to electric $p$-branes and
forms
\beq{1.10}
{\cal F}^s= Q_s \tau(\bar I_s),
\eeq
correspond to magnetic $p$-branes; $Q_s \neq 0$, $s \in S$.
In (\ref{1.10})  and in what follows
\beq{1.13a}
\bar I\equiv\{1,\ldots,n\}\setminus I.
\eeq

All the  manifolds $M_{i}$, $i > 1$, are assumed to be oriented and
connected and  the volume $d_i$-forms
\beq{1.12}
\tau_i  \equiv \sqrt{|g^i(y_i)|}
\ dy_i^{1} \wedge \ldots \wedge dy_i^{d_i},
\eeq
are well--defined for all $i=  1,\ldots,n$.
Here $d_{i} =   {\rm dim} M_{i}$, $i =   1, \ldots, n$
(in spherically symmetric case $M_1 = S^{d_1}$), $d_1 > 1$,
$D =   1 + \sum_{i =   1}^{n} d_{i}$, and for any
 $I =   \{ i_1, \ldots, i_k \} \in \Omega$, $i_1 < \ldots < i_k$,
we denote
\ber{1.13}
\tau(I) \equiv \tau_{i_1}  \wedge \ldots \wedge \tau_{i_k},
\\ \label{1.14}
M_{I} \equiv M_{i_1}  \times  \ldots \times M_{i_k}, \\
\label{1.15}
d(I) \equiv {\rm dim } M_I =  \sum_{i \in I} d_i.
\eer

The parameters  $h_s$ appearing in the solution
satisfy the relations
\beq{1.16}
h_s = K_s^{-1}, \qquad  K_s = B_{s s},
\eeq
where
\ber{1.17}
B_{ss'} \equiv
d(I_s\cap I_{s'})+\frac{d(I_s)d(I_{s'})}{2-D}+
\chi_s\chi_{s'}\lambda_{\alpha a_s}\lambda_{\beta a_{s'}}
h^{\alpha\beta},
\eer
$s, s' \in S$, with $(h^{\alpha\beta})=(h_{\alpha\beta})^{-1}$.
Here we assume that
\beq{1.17a}
({\bf i}) \qquad B_{ss} \neq 0,
\eeq
for all $s \in S$, and
\beq{1.18b}
({\bf ii}) \qquad {\rm det}(B_{s s'}) \neq 0,
\eeq
i.e. the matrix $(B_{ss'})$ is a non-degenerate one. In (\ref{1.9})
another non-degenerate matrix (``a quasi-Cartan'' matrix)
appears
\beq{1.18}
(A_{ss'}) = \left( 2 B_{s s'}/B_{s' s'} \right).
\eeq
Here  some ordering in $S$ is assumed.

This matrix also appears in  the relations for
\beq{1.19}
f_s = \exp( - q^s),
\eeq
where $(q^s) = (q^s(u))$ is a solution to Toda-type equations
\beq{1.20}
\ddot{q^s} = -  B_s \exp( \sum_{s' \in S} A_{s s'} q^{s'} ),
\eeq
with
\beq{1.21}
 B_s = 2 K_s A_s,  \quad  A_s =  \frac12  \eps_s Q_s^2,
\eeq
$s \in S$. Here
\beq{1.22}
\eps_s=(-\eps[g])^{(1-\chi_s)/2}\eps(I_s) \theta_{a_s},
\eeq
$s\in S$, $\eps[g]\equiv\sign\det(g_{MN})$. More explicitly
(\ref{1.22}) reads: $\eps_s=\eps(I_s) \theta_{a_s}$ for
$v_s = e$ and $\eps_s=-\eps[g] \eps(I_s) \theta_{a_s}$, for
$v_s = m$.

In (\ref{1.3})
\bear{1.23}
f_1(u) =R \sh(\sqrt{C_1}u), \ C_1>0, \ \xi_1 w>0;
\\ \label{1.24}
R \sin(\sqrt{|C_1|}u), \ C_1<0, \  \xi_1 w>0;   \\ \label{1.25}
R \ch(\sqrt{C_1}u),  \ C_1>0, \ \xi_1w <0; \\ \label{1.26}
\left|\xi_1(d_1-1)\right|^{1/2} u, \ C_1=0,  \ \xi_1w>0,
\ear
where  $C_1$ is constant and $R =  |\xi_1(d_1-1)/C_1|^{1/2}$.

Vectors $c=(c^A)= (c^i, c^\alpha)$ and
$\bar c=(\bar c^A)$ satisfy the linear constraints
\bear{1.27}
U^s(c)= \sum_{i \in I_s}d_ic^i-\chi_s\lambda_{a_s\alpha}c^\alpha=0,
\\ \label{1.28}
U^s(\bar c)=  \sum_{i\in I_s}d_i\bar c^i-
\chi_s\lambda_{a_s\alpha}\bar c^\alpha=0,
\ear
$s\in S$,
\bear{1.29}
U^1( c) =  -c^1+\sum_{j=1}^n d_j c^j=0, \\ \label{1.30}
U^1(\bar c) =
-\bar c^1+\sum_{j=1}^nd_j\bar c^j=0,
\ear
and
\beq{1.30a}
C_1 \frac{d_1}{d_1-1}= 2 E_{TL} +
h_{\alpha\beta}c^\alpha c^\beta+ \sum_{i=2}^nd_i(c^i)^2+
\frac1{d_1-1}\left(\sum_{i=2}^nd_ic^i\right)^2,
\eeq
where
\beq{1.31}
E_{TL} = \frac{1}{4}  \sum_{s,s' \in S} h_s
A_{s s'} \dot{q^s} \dot{q^{s'}}
  + \sum_{s \in S} A_s  \exp( \sum_{s' \in S} A_{s s'} q^{s'} ),
\eeq
is an integration constant (energy) for the solutions from
(\ref{1.20}).

We note that the eqs. (\ref{1.20}) correspond to the
Toda-type Lagrangian
\beq{1.31a}
L_{TL} = \frac{1}{4}  \sum_{s,s' \in S}
h_s  A_{s s'} \dot{q^s}\dot{q^{s'}}
-  \sum_{s \in S} A_s  \exp( \sum_{s' \in S} A_{s s'} q^{s'} ).
\eeq

{\bf Remark 1.}
{\em Here we identify notations  for $g^{i}$  and  $\hat{g}^{i}$, where
$\hat{g}^{i} = p_{i}^{*} g^{i}$ is the
pullback of the metric $g^{i}$  to the manifold  $M$ by the
canonical projection: $p_{i} : M \rightarrow  M_{i}$, $i = 1,
\ldots, n$. An analogous agreement will be also kept for volume forms etc.}

Due to (\ref{1.9}) and  (\ref{1.10}), the dimension of
$p$-brane worldsheet $d(I_s)$ is defined by
\ber{1.16a}
d(I_s)=  n_{a_s}-1, \quad d(I_s)=   D- n_{a_s} -1,
\eer
for $s \in S_e, S_m$ respectively.
For a $p$-brane: $p =   p_s =   d(I_s)-1$.

The solutions are valid if the following  restrictions on the sets
$\Omega_{a,v}$ are imposed.
These restrictions guarantee the block-diagonal structure
of the stress-energy tensor, like for the metric, and the existence of
$\sigma$-model representation \cite{IMC} (see also \cite{AR}).
We denote $w_1\equiv\{i|i\in \{2,\dots,n\},\quad d_i=1\}$, and
$n_1=|w_1|$ (i.e. $n_1$ is the number of 1-dimensional spaces among
$M_i$, $i=1,\dots,n$).

{\bf Restriction 1.} {\em Let 1a) $n_1\le1$ or 1b) $n_1\ge2$ and for
any $a\in\tri$, $v\in\{e,m\}$, $i,j\in w_1$, $i<j$, there are no
$I,J\in\Omega_{a,v}$ such that $i\in I$, $j\in J$ and $I\setminus\{i\}=
J\setminus\{j\}$.}

{\bf Restriction 2.} {\em Let 2a) $n_1=0$ or
2b) $n_1\ge1$ and for any $a\in\tri$, $i\in w_1$ there are no
$I\in\Omega_{a,m}$, $J\in\Omega_{a,e}$ such that $\bar I=\{i\}\sqcup J$.}

These restrictions are  satisfied in the non-composite case
\cite{IM0,IM}:  $|\Omega_{a,e}| + |\Omega_{a,m}| = 1$,
(i.e when there are no two  $p$-branes with the same color index $a$,
$a\in\tri$.) Restriction 1 and 2 forbid certain intersections of two
$p$-branes with the same color index for  $n_1 \geq 2$ and  $n_1 \geq 1$
respectively.  Restriction 2 is satisfied identically if all
$p$-branes contain a common manifold $M_j$ (say, time manifold).

This solution describes a set of charged (by forms) overlapping
$p$-branes ($p_s=d(I_s)-1$, $s \in S$) ``living'' on submanifolds
of $M_2 \times \dots \times M_n$.

\subsection{$U^s$-vectors and scalar products}

Here we consider a minisuperspace
covariant form of constraints
and corresponding scalar products
that will be used in the next section.
The linear constraints (\ref{1.27})-(\ref{1.30})
may be written in the following form
\beq{1.32}
U^r(c)= U^r_A c^A= 0, \qquad U^r(\bar c)= U^r_A \bar c^A= 0,
\eeq
$r = s,1$, where
\ber{1.33}
(U_A^s)=(d_i\delta_{iI_s},-\chi_s\lambda_{\alpha a_s}),
\eer
$s=(a_s,v_s,I_s) \in S$, and
\ber{1.34}
(U_A^1)=(- \delta^1_i + d_i, 0),
\eer
$A = (i, \alpha)$.

The quadratic constraint (\ref{1.30a}) reads
\beq{1.30b}
E=E_1+ E_{TL}+ \frac12 \hat G_{AB} c^A c^B = 0,
\eeq
where $C_1=2E_1(U^1,U^1)$,
\beq{1.35}
(U^1,U^1) = 1/d_1 - 1,
\eeq
($d_1 > 1$) and
\ber{1.36}
(\hat G_{AB})=\barr{cc}
G_{ij}& 0\\
0& h_{\alpha\beta}
\earr,
\eer
is the target space metric with
\ber{1.37}
G_{ij}= d_i \delta_{ij} - d_i d_j,
\eer
$i,j = 1, \ldots, n$.
In (\ref{1.35})  a scalar product appears
\ber{1.38}
(U,U')=\hat G^{AB}U_AU'_B,
\eer
where $U=U_Az^A$, $U' = U_A z^A$ are linear functions
on $\R^{n+l}$, and  $(\hat G^{AB})=(\hat G_{AB})^{-1}$.
The scalar products (\ref{1.38}) for co-vectors
$U^s$ from $(\ref{1.33})$  were calculated in
\cite{IMC}
\beq{1.39}
(U^s,U^{s'})= B_{s s'},
\eeq
$s, s' \in S$ (see (\ref{1.17})).
It follows from (\ref{1.18b}) and  (\ref{1.39})
that the vectors $U^s$, $s \in S$, are  linearly
independent. Hence, the number of the vectors
$U^s$ should not exceed the dimension of
the dual space $({\bf R}^{n+ l})^{*}$, i.e.
\beq{1.40c}
|S| \leq n+ l.
\eeq
We also get \cite{IMC}
\beq{1.39a}
(U^s,U^{1})= 0,
\eeq
for all $s \in S$. This relation takes place, since
all $p$-branes do not live in $M_1$: $I_s \in \{2,\ldots,n \}$.

{\bf Intersection rules.}
>From  (\ref{1.16}), (\ref{1.17}) and (\ref{1.18})  we get
the  intersection rules  corresponding
to the quasi-Cartan matrix $(A_{s s'})$ \cite{IMJ}
\beq{1.40}
d(I_s \cap I_{s'})= \frac{d(I_s)d(I_{s'})}{D-2}-
\chi_s\chi_{s'}\lambda_{a_s}\cdot\lambda_{a_{s'}} + \frac12 K_{s'} A_{s s'},
\eeq
where $\lambda_{a_s}\cdot\lambda_{a_{s'}} =
\lambda_{\alpha a_s}\lambda_{\beta a_{s'}} h^{\alpha\beta}$,
$s, s' \in S$.

The contravariant components $U^{rA}= \hat G^{AB} U^r_B$ reads
\cite{IMC,IMJ}
\beq{1.41}
U^{si}= G^{ij}U_j^s=
\delta_{iI_s}-\frac{d(I_s)}{D-2}, \quad U^{s\alpha}= - \chi_s
\lambda_{a_s}^\alpha,
\eeq
\beq{1.42}
U^{1i}=-\frac{\delta_1^i}{d_1},
\quad U^{1\alpha}=0,
\eeq
$s \in S$.
Here (as in \cite{IMZ})
\beq{1.43}
G^{ij}=\frac{\delta^{ij}}{d_i}+\frac1{2-D},
\eeq
$i,j=1,\dots,n$, are the components of the matrix inverse to
$(G_{ij})$ from (\ref{1.37}). The contravariant components
(\ref{1.41}) and  (\ref{1.42}) occur as powers in relations
for the metric and scalar fields in (\ref{1.3}) and  (\ref{1.4}).

We note that the solution under consideration in a special case
of $A_m$ Toda chain was obtained earlier  in \cite{GM1}.
Special $A_1 \oplus \dots \oplus A_1$ Toda case, when vectors
$U^s$ are mutually orthogonal, was considered earlier in \cite{IMJ}
(for non-composite case see also \cite{BGIM,GrIM,BIM}).
For a (general) block-orthogonal set of vectors $U^s$
special solutions were considered in \cite{Br,IMJ2}.

\section{Spherically symmetric solutions. Black holes.}

\subsection{The choice of parameters }

Here we consider the spherically symmetric case:
\beq{2.1}
w = 1, \quad M_1 = S^{d_1}, \quad g^1 = d \Omega^2_{d_1},
\eeq
where $d \Omega^2_{d_1}$ is the canonical metric on a unit sphere
$S^{d_1}$, $d_1 \geq 2$. In this case $\xi^1 = d_1 -1$.
We also assume that
\beq{2.2}
M_2 = \R, \qquad g^2 = - dt \otimes dt,
\eeq
i.e.  $M_2$ is a time manifold.

We put $C_1 \geq 0$.
In this case relations (\ref{1.23})-(\ref{1.26}) read
\bear{1.23s}
f_1(u) = \bar d \frac{\sh(\sqrt{C_1}u)}{\sqrt{C_1}},
\ C_1>0,
\\ \label{1.26s}
\bar du, \ C_1=0.
\ear
Here and in what follows
\beq{2.3}
\bar d = d_1-1.
\eeq

Let us consider the null-geodesic equations
for the light ``moving'' in the radial direction
(following from $ds^2 =0$):
\bear{2.4}
 \frac{dt}{du} = \pm \Phi, \\ \label{2.4a}
 \Phi =  f_1^{d_1/(1-d_1)} e^{(c^1 - c^2) u +  \bar c^1 - \bar c^2}
\prod_{s\in S} f_s^{- 2 h_s  \delta_{2 I_s}},
\ear
equivalent to
\beq{2.5}
 t - t_0 = \pm \int_{u_0}^{u} d \bar u  \Phi(\bar u),
\eeq
where $t_0, u_0$ are constants.

Let us consider   solutions
(defined on some interval $[u_0, +\infty)$) with a  horizon
at $u = + \infty$ satisfying
\beq{2.6}
  \int_{u_0}^{ + \infty} d  u  \Phi( u) = + \infty.
\eeq

Here we restrict ourselves to  solutions with
$C_1 > 0$ and linear asymptotics at infinity
\beq{2.7}
q^s = - \beta^s u + \bar \beta^s  + o(1),
\eeq
$u \to +\infty$, where $\beta^s, \bar \beta^s$ are
constants, $s \in S$. This relation gives us an
asymptotical solution to  Toda type eqs. (\ref{1.20}) if
\beq{2.8}
\sum_{s' \in S} A_{s s'} \beta^{s'} > 0,
\eeq
for all $s \in S$. In this case the energy  (\ref{1.31})
reads
\beq{2.9}
E_{TL} = \frac{1}{4}  \sum_{s,s' \in S}
h_s A_{s s'}  \beta^s \beta^{s'}.
\eeq

{\bf Remark 2.} {\em  For positive-definite matrices
$(h_s A_{s s'})$ and  $(h_{\alpha\beta})$ we get from
(\ref{1.30a}) and (\ref{2.9}):
$E_{TL} \geq 0$,  $C_{1} \geq 0$.
(For the extremal case $E_{TL} = C_{1} = 0$ see Sect. 7.)
According to  Lemma 2 from \cite{Br2}
black hole solutions can only exist for
$C_{1} \geq 0$ and the horizon is then at
$u = \infty$.}

For  the function  (\ref{2.4a}) we get
\beq{2.10}
\Phi(u) \sim \Phi_0 e^{\beta u}, \quad u \to +\infty,
\eeq
where  $\Phi_0 \neq 0$ is  constant,
\beq{2.11}
 \beta = c^1 - c^2 + \sqrt{C_1} h_1
+ \sum_{s \in S} \beta_s h_s  \delta_{2 I_s},
\eeq
and
\beq{2.12}
  h_1 = (U^1, U^1)^{-1} = \frac{d_1}{1 - d_1}.
\eeq
Horizon at $u = + \infty$  takes place if and only if
\beq{2.12a}
  \beta \geq 0.
\eeq
Let us introduce dimensionless parameters
\beq{2.13}
b^s = \beta^s / \sqrt{C_1}, \qquad
b^A =  c^A / \sqrt{C_1},
\eeq
where $s \in S$, $A = (i, \alpha)$,
$C_1 > 0$.

Thus, a horizon at $u = + \infty$
corresponds to a point $b = (b^s,b^A) \in  \R^{|S| + n +l}$
satisfying the relations following from
(\ref{1.32}), (\ref{1.30b}), (\ref{2.8}), (\ref{2.9})
and (\ref{2.11})-(\ref{2.13}):
\bear{2.14}
U^r_A b^A= 0,   \qquad r = s,1; \ s \in S, \\ \label{2.15}
\frac{1}{2}  \sum_{s,s' \in S} h_s A_{s s'}  b^s b^{s'}
+ \hat G_{AB} b^A b^B = |h_1|, \\
\label{2.16}
\sum_{s' \in S} A_{s s'} b^{s'} > 0,  \\ \label{2.17}
 f(b) \equiv
 b^1 - b^2  + \sum_{s\in S} b_s h_s  \delta_{2 I_s} \geq |h_1|.
\ear

{\bf Proposition 1.} {\em Let  matrix $(h_{\alpha\beta})$
be positively defined. Then the point $b = (b^s,b^A)$ satisfying
relations (\ref{2.14})-(\ref{2.17}) exists only if
\beq{2.18}
2 \in I_s, \quad \forall s \in S,
\eeq
(i. e. all p-branes have a common time direction $t$)
and is unique: $b = b_0$, where
\bear{2.19}
b_0^A  = - \delta^{A}_{2} + h_1 U^{1 A}  +
\sum_{s\in S}  h_s b_0^s U^{s A},  \\ \label{2.20}
b_0^s = 2 \sum_{s' \in S} A^{s s'},
\ear
where $s \in S$, $A = (i, \alpha)$,
and the matrix $(A^{s s'})$ is inverse to the matrix
$(A_{s s'}) = ( 2 (U^s, U^{s'})/ (U^{s'}, U^{s'}))$. }

{\bf Proof.} Let ${\cal E}$ be a manifold described
by relations (\ref{2.14})-(\ref{2.15}). This manifold is an ellipsoid.
Indeed, due to positively definiteness of $(h_{\alpha\beta})$
the matrix $\hat G_{AB}$ has a signature $(-,+, \ldots,+)$,
since the matrix $(G_{ij})$ from (\ref{1.37}) has a signature
$(-,+, \ldots,+)$ \cite{IMZ}. Due to relations
$(U^1,U^1) < 0$,  $(U^1,U^s) = 0$, $(U^s,U^s) \neq  0$ for all $s \in S$,
and (\ref{1.18b}) the matrices $(B_{s s'})$ and $(A_{s s'})$ are positively
defined and all $h_s > 0$, $s \in S$. Then, the quadratic form
in (\ref{2.15}) has a pseudo-Euclidean signature.
Due to $(U^1,U^1) < 0$ the intersection of the
hyperboloid (\ref{2.15}) with the (multidimensional)
plane $U^1_A z^A = 0$ gives us an ellipsoid. Its intersection
with the planes $U^s_A z^A = 0$, $s \in S$, give us to an ellipsoid,
coinciding with ${\cal E}$.

Let us consider  a function $f_{|} : {\cal E} \rightarrow \R$
that is a restriction of the linear function
(\ref{2.17}) on ${\cal E}$.
Let $b_{*} \in {\cal E}$ be a point of maximum
of $f_{|}$.
Using the conditional extremum method
and the  fact that ${\cal E}$ is  ellipsoid
we  prove that

that
\bear{2.21}
b_{*}^A = - \delta^{A}_{2} + h_1 U^{1 A} +
\sum_{s\in S}  h_s b_{*}^s U^{s A},  \\ \label{2.21a}
b_{*}^s = 2 \sum_{s' \in S} A^{s s'} \delta_{2 I_{s'}},
\ear
$s \in S$, $A = (i, \alpha)$.
Let us consider the function
\bear{2.22}
 \bar f(b, \lambda) \equiv  f(b)
 - \lambda_1 U^1_A b^A
 - \sum_{s \in S} \lambda_s U^s_A b^A
 - \lambda_0
\left(\sum_{s,s' \in S} \frac{h_s}{2} A_{s s'} b^s b^{s'}
+ \hat G_{AB} b^A b^B + h_1 \right),
\ear
where
$\lambda = (\lambda_0, \lambda_1, \lambda_s)$ is a vector of
Lagrange multipliers. The  points of extremum for the
function  $\bar f$ from (\ref{2.22}) have the
form $(\lambda_0  b_{*}, \lambda)$ with $b_{*}$ from (\ref{2.21}) and
\bear{2.23}
  \lambda_0 = \pm 1, \quad   \lambda_1 = 1/(d_1 -1), \quad
  \lambda_s = - 2 \sum_{s' \in S} h_s A^{s s'} \delta_{2 I_{s'}},
\ear
$s \in S$. Then, the  points $b_{*}$ and $ - b_{*}$
are the points of maximum
and minimum, respectively, for the function $f_{|}$ defined on the
ellipsoid  ${\cal E}$. Since $f(b_{*}) = |h_1|$, the only point
satisfying the restriction $f(b) \geq |h_1|$ is $b = b_{*}$.
>From (\ref{2.16}) we get
\bear{2.24}
\sum_{s' \in S} A_{s s'} b^{s'}= 2 \delta_{2 I_{s}} > 0
\Longleftrightarrow  2 \in I_s,
\ear
for all $s \in S$. The proposition is proved.

We introduce a new radial variable $R = R(u)$ by relations
\bear{2.28}
\exp( - 2\bar{\mu} u) = 1 - \frac{2\mu}{R^{\bar{d}}} = F,
\qquad \bar{\mu} = \sqrt{C_1}, \quad
\mu = \bar{\mu}/ \bar{d} >0,
\ear
$u > 0$, $R^{\bar d} > 2\mu$ ($\bar d = d_1 -1$).
We put
\bear{2.27}
\bar{c}^A = 0,
\\     \label{2.27f}
q^s(0) = 0.
\ear
$A = (i, \alpha)$, $s \in S$.
These relations guarantee the asymptotical flatness
(for $R \to +\infty$) of the $(2+d_1)$-dimensional section of the metric.

Let us denote
\beq{2.28a}
H_s = f_s e^{- \bar{\mu} b^s_0 u },
\eeq
$s \in S$.
Then,  solutions (\ref{1.3})-(\ref{1.5}) may be written as follows
\bear{2.30}
g= \Bigl(\prod_{s \in S} H_s^{2 h_s d(I_s)/(D-2)} \Bigr)
\biggl\{ F^{-1} dR \otimes dR
+ R^2  d \Omega^2_{d_1}  \\ \nn
-  \Bigl(\prod_{s \in S} H_s^{-2 h_s} \Bigr) F  dt \otimes dt
+ \sum_{i = 3}^{n} \Bigl(\prod_{s\in S}
  H_s^{-2 h_s \delta_{iI_s}} \Bigr) g^i  \biggr\},
\\  \label{2.31}
\exp(\varphi^\alpha)=
\prod_{s\in S} H_s^{h_s \chi_s \lambda_{a_s}^\alpha},
\\  \label{2.32a}
F^a= \sum_{s \in S} \delta^a_{a_s} {\cal F}^{s},
\ear
where
\beq{2.32}
{\cal F}^s= - \frac{Q_s}{R^{d_1}}
\left( \prod_{s' \in S}  H_{s'}^{- A_{s s'}} \right) dR \wedge\tau(I_s),
\eeq
$s\in S_e$,
\beq{2.33}
{\cal F}^s= Q_s \tau(\bar I_s),
\eeq
$s\in S_m$.
Here $Q_s \neq 0$, $h_s =K_s^{-1}$; parameters $K_s \neq 0$ and
the non-degenerate
matrix $(A_{s s'})$ are defined by  relations (\ref{1.40})
and $(A_{s s}) = 2$, $s \in S$.

Functions $H_s > 0$ obey the equations
\beq{2.34}
 R^{d_1} \frac{d}{dR} \left( R^{d_1}
\frac{F}{H_s}	\frac{d H_s}{dR} \right) = B_s
\prod_{s' \in S}  H_{s'}^{- A_{s s'}},
\eeq
$s \in S$, where $B_s \neq 0$ are defined
in (\ref{1.21}) and (\ref{1.22}).
These equations follow from Toda-type equations (\ref{1.20}) and
the definition   (\ref{2.28}) and   (\ref{2.28a}).

It follows from (\ref{2.7}), (\ref{2.13}) and  (\ref{2.28a})
that there exist finite limits
\beq{2.35a}
H_s  \to H_{s0} \neq 0,
\eeq
for $R^{\bar d} \to 2\mu$, $s \in S$.
We note, that in this case the metric (\ref{2.30})
does really have a horizon at  $R^{\bar{d}} =   2 \mu$.

>From (\ref{2.27f})  we get.
\beq{2.35}
H_s (R = +\infty) = 1,
\eeq
$s \in S$,

The metric (\ref{2.30}) has a regular horizon at
$R^{\bar{d}} =   2 \mu$.
The Hawking temperature corresponding to
the solution is (see also \cite{Oh,BIM} for orthogonal case)
found to be
 \beq{2.36}
T_H=   \frac{\bar{d}}{4 \pi (2 \mu)^{1/\bar{d}}}
\prod_{s \in S} H_{s0}^{- h_s},
\eeq
where $H_{s0}$ are defined in (\ref{2.35a})

The boundary conditions (\ref{2.35a}) and  (\ref{2.35})
play a crucial role here, since they
single out, generally speaking, only few solutions
to eqs. (\ref{2.34}).

Moreover for some values of parameters $\mu = \bar \mu / \bar d$, $\eps_s$
and $Q_s^2$ the solutions to eqs. (\ref{2.34})-(\ref{2.35})
do not exist. Indeed, from (\ref{1.21}), (\ref{1.31}),
(\ref{2.9}), (\ref{2.13}), (\ref{2.20}), (\ref{2.28}) and
(\ref{2.27f} we get
\beq{2.35b}
E_{TL} = \bar \mu^2  \sum_{s,s' \in S} h_s A^{ss'} =
\frac{1}{4}  \sum_{s,s' \in S} h_s A_{s s'}
\dot{q^s}(0) \dot{q^{s'}}(0) + \sum_{s \in S}
\frac12  \eps_s Q_s^2.
\eeq
Let the matrix $(h_s A_{s s'})$ be positive-definite
(in this case matrix $(B_{s s'})$ is positive-definite
too and all $h_s > 0$). Then $E_{TL} > 0$ and
\beq{2.35c}
\bar \mu^2  \sum_{s,s' \in S} h_s A^{ss'} \geq
\sum_{s \in S}   \frac12  \eps_s Q_s^2.
\eeq
If the parameters  obey the relation
\beq{2.35d}
0 < \bar \mu^2  \sum_{s,s' \in S} h_s A^{ss'} <
\sum_{s \in S}  \frac12  \eps_s Q_s^2,
\eeq
e.g. for $\eps_s = + 1$ and big enough $Q_s^2$, the
solution under consideration does not exist.

We note that the solution to eqs.
(\ref{2.34})-(\ref{2.35}) may not be unique.
The simplest example occurs in the case of one $p$-brane,
when $h_s >0$, $\eps_s = +1$ and $\bar \mu^2 h_s > Q_s^2$.
In this case we have two solutions to (\ref{2.34})-(\ref{2.35})
corresponding to two possible values of $\dot{q^s}(0)$.


Thus, we obtained a family of black hole solutions
up to solutions of radial equations (\ref{2.34})
with the boundary conditions (\ref{2.35a}) and  (\ref{2.35}).
In the next sections we consider several exact solutions
to eqs. (\ref{2.34})-(\ref{2.35}).

{\bf Remark 3.} {\em Let $M_i = \R$ and $g^i = -d \bar t \otimes d\bar t$
for some $i \geq 3$. Then the metric (\ref{2.30}) has no a horizon
with respect to the ``second time''  $\bar t$ for $R^{\bar{d}} \to 2\mu$.
Thus, we a led to a ``single-time'' theorem from \cite{Br2}.
Relation (\ref{2.18}) from Proposition 1 coincides with
the ``no-hair'' theorem from \cite{Br2}. }

\section{Polynomial structure of $H_s$ for  Lie algebras}

\subsection{Conjecture on polynomial structure}

Now we deal  with solutions to second order non-linear
differential equations  (\ref{2.34}) that may be rewritten
as follows
\beq{3.1}
 \frac{d}{dz} \left( \frac{F}{H_s}	\frac{d}{dz} H_s \right) = \bar B_s
\prod_{s' \in S}  H_{s'}^{- A_{s s'}},
\eeq
where $H_s(z) > 0$, $F =  1 - 2\mu z$, $\mu > 0$,
$z = R^{-\bar d}$, $\bar B_s =  B_s/ \bar d^2 \neq 0$.
Eqs. (\ref{2.35}) and  (\ref{2.35a}) read
\bear{3.2a}
H_{s}((2\mu)^{-1} -0) = H_{s0} \in (0, + \infty), \\
\label{3.2b}
H_{s}(0) = 1,
\ear
$s \in S$.

It seems rather difficult to find the solutions to a set
of eqs. (\ref{3.1})-(\ref{3.2b}) for arbitrary
values of parameters $\mu$, $\bar B_s$, $s \in S$ and
quasi-Cartan matrices $A =(A_{s s'})$. But we may
expect a drastically simplification of the problem
under consideration for certain class of parameters and/or
$A$-matrices.

In general we may try to seek solutions of (\ref{3.1})  in a class
of functions analytical in a disc $|z| < L$ and continuous
in semi-interval $0 < z \leq (2\mu)^{-1}$. For $|z| < L$
we get
\beq{3.3}
H_{s}(z) = 1 + \sum_{k = 1}^{\infty} P_s^{(k)} z^k,
\eeq
where $P_s^{(k)}$ are constants, $s \in S$. Substitution
of (\ref{3.3})  into (\ref{3.1}) gives us an infinite
chain of relations on parameters $P_s^{(k)}$  and
$\bar B_s$.  In general case it seems to be impossible
to solve this chain of equations.

Meanwhile there exist solutions to eqs. (\ref{3.1})-(\ref{3.2b})
of polynomial type. The simplest example occurs in orthogonal
case \cite{CT,AIV,Oh,IMJ,BIM}, when
\beq{3.4}
(U^s,U^{s'})= B_{s s'} = 0,
\eeq
for  $s \neq s'$, $s, s' \in S$. In this case
$(A_{s s'}) = {\rm diag}(2,\ldots,2)$ is a Cartan matrix
for semisimple Lie algebra $A_1 \oplus  \ldots  \oplus  A_1$
and
\beq{3.5}
H_{s}(z) = 1 + P_s z,
\eeq
with $P_s \neq 0$, satisfying
\beq{3.5a}
P_s(P_s + 2\mu) = -\bar B_s,
\eeq
$s \in S$.

In \cite{Br,IMJ2,CIM} this solution
was generalized to a block orthogonal
case:
\ber{3.6}
S=S_1 \cup\dots\cup S_k, \qquad  S_i \cap S_j = \emptyset, \quad i \neq j,
\eer
$S_i \ne \emptyset$, i.e. the set $S$ is a union of $k$ non-intersecting
(non-empty) subsets $S_1,\dots,S_k$,
and
\ber{3.7}
(U^s,U^{s'})=0
\eer
for all $s\in S_i$, $s'\in S_j$, $i\ne j$; $i,j=1,\dots,k$.
In this case (\ref{3.5}) is modified as follows
\beq{3.8}
H_{s}(z) = (1 + P_s z)^{b_0^s},
\eeq
where $b_0^s$ are defined in  (\ref{2.20})
and parameters $P_s$  and are coinciding inside
blocks, i.e. $P_s = P_{s'}$ for $s, s' \in S_i$, $i =1,\dots,k$.
Parameters $P_s \neq 0 $ satisfy the relations (\ref{3.5a})
and parameters $\bar B_s$  are also  coinciding inside
blocks, i.e. $\bar B_s = \bar B_{s'}$
for $s, s' \in S_i$, $i =1,\dots,k$. In this case $H_s$
are analytical in $|z| < L$, where $L =
{\rm min} (| P_ s|^{-1}, s \in S$).

Let $(A_{s s'})$ be  a Cartan matrix  for a  finite-dimensional
semisimple Lie  algebra $\cal G$. In this case all powers in
(\ref{2.20})  are  natural numbers  \cite{GrI}
\beq{3.11}
b_0^s = 2 \sum_{s' \in S} A^{s s'} = n_s \in \N,
\eeq
and  hence, all functions $H_s$ are polynomials, $s \in S$.

{\bf Conjecture 1.} {\em Let $(A_{s s'})$ be  a Cartan matrix
for a  semisimple finite-dimensional Lie algebra $\cal G$.
Then  the solution to eqs. (\ref{3.1})-(\ref{3.2b})
(if exists) is a polynomial
\beq{3.12}
H_{s}(z) = 1 + \sum_{k = 1}^{n_s} P_s^{(k)} z^k,
\eeq
where $P_s^{(k)}$ are constants,
$k = 1,\ldots, n_s$, integers $n_s = b_0^s$ are
defined in (\ref{3.11}) and $P_s^{(n_s)} \neq 0$,  $s \in S$}.

In extremal case ($\mu = + 0$) an a analogue of this conjecture
was suggested previously in \cite{LMMP}.

\subsection{Proof of Conjecture 1 for $A_m$ and $C_{m+1}$ }

First, we prove the Conjecture for
simple Lie algebras $A_{m}= sl(m+1)$, $m \geq 1$.
Let us  consider exact solutions to equations of motion
of a Toda-chain corresponding to the Lie algebra
 $A_{m}$ \cite{T,And} ,
\beq{B.1}
\ddot q^s  = - B_s \exp\left( \sum_{s'=1}^{m} A_{s s'} q^{s'}  \right) ,
\eeq
where
\beq{B.1a}
\left(A_{ss'}\right)=
\left( \begin{array}{*{6}{c}}
2&-1&0&\ldots&0&0\\
-1&2&-1&\ldots&0&0\\
0&-1&2&\ldots&0&0\\
\multicolumn{6}{c}{\dotfill}\\
0&0&0&\ldots&2&-1\\
0&0&0&\ldots&-1&2
\end{array}
\right)\quad
\eeq
is the Cartan matrix  of the Lie algebra $A_{m}$ and $B_s > 0$,
$s,s' = 1, \ldots, m$.  Here we put
$S =  \{1, \ldots, m \}$.

The equations of motion   (\ref{B.1}) correspond
to the Lagrangian
\beq{B.2}
 L_T = \frac{1}{2} \sum_{s,s'=1}^{m} A_{ss'} \dot q^s  \dot q^{s'}  -
\sum_{s=1}^{m}  B_s \exp \left( \sum_{s'=1}^{m} A_{ss'} q^{s'}  \right).
\eeq
This Lagrangian may be obtained from the standard one \cite{T}
by separating a coordinate describing the  motion  of the center of mass.

Using the result of A. Anderson \cite{And}
we present the solution to eqs. (\ref{B.1}) in the following form
\beq{B.3}
C_s \exp(-q^s(u)) =
\sum_{r_1< \dots <r_s}^{m+1} v_{r_1}\cdots v_{r_s}
\Delta^2( w_{r_1}, \ldots, w_{r_s}) \exp[(w_{r_1}+\ldots +w_{r_s})u],
\eeq
$s = 1, \ldots, m$, where
\beq{B.4a}
\Delta( w_{r_1}, \ldots, w_{r_s})  =
\prod_{i<j}^{s} \left(w_{r_i}-w_{r_j}\right); \quad
\Delta(w_{r_1}) \equiv 1,
\eeq
denotes the  Vandermonde determinant.
The real constants $v_r$ and $w_r$, $r = 1, \ldots, m + 1$, obey the
relations
\beq{B.5}
\prod_{r=1}^{m+1} v_r= \Delta^{-2}(w_1,\ldots, w_{m+1}), \qquad
\sum_{r=1}^{m+1}w_r=0.
\eeq
In (\ref{B.3})
\beq{B.6}
C_s = \prod_{s'=1}^{m} B_{s'}^{-A^{s s'}},
\eeq
where
\ber{B.7}
A^{s s'}= \frac1{m+1}\min(s,s')[m+1 - \max(s,s')],
\eer
$s, s' = 1, \ldots, m$,  are components of a matrix inverse to the Cartan one, i.e.
$(A^{s s'})=(A_{ss'})^{-1}$ (see Ch.7 in  \cite{FS}).
Here
\beq{B.8}
v_r \neq 0, \qquad w_r \neq w_{r'}, \quad r \neq r',
\eeq
$r, r' = 1, \ldots, m +1$.
We note that the solution with $B_s > 0$ may be obtained
>from the solution  with $B_s =1$ (see \cite{And}) by a certain shift $q^s
\mapsto q^s + \delta^s$.

The energy reads \cite{And}
\beq{B.9}
 E_T= \frac{1}{2} \sum_{s,s'=1}^{m} A_{s s'} \dot q^s  \dot q^{s'} +
\sum_{s=1}^{m}  B_s \exp \left( \sum_{s'=1}^{m} A_{ss'} q^{s'}  \right) =
\frac{1}{2}\sum_{r=1}^{m+1} w^2_r.
\eeq

If  $B_s > 0$, $s \in S$, then  all $w_r, v_r$ are real and, moreover, all
$v_r > 0$, $r =1, \ldots, m+1$.  In a general case $B_s \neq 0$, $s \in
S$, relations  (\ref{B.3})-(\ref{B.6}) also describe real solutions to
eqs. (\ref{B.1}) for suitably chosen complex
parameters $v_r$ and $w_r$. These parameters are either real or belong to
pairs of complex conjugate (non-equal) numbers, i.e., for example, $w_1 =
\bar w_2$, $v_1 = \bar v_2 $. When some of $B_s$ are negative, there are
also some special (degenerate) solutions to eqs. (\ref{B.1}) that are not
described by relations (\ref{B.3})-(\ref{B.6}), but may be obtained from
the latter by certain limits of parameters $w_r$.

For the energy (\ref{1.31}) we get
\beq{B.10}
E_{TL} = \frac{1}{2K} E_T=  \frac{h}{4} \sum_{r=1}^{m+1} w^2_r.
\eeq
Here
\beq{B.11}
K_s = K,    \qquad  h_s= h = K^{-1},
\eeq
$s \in S$.
Thus, in the $A_m$ Toda chain case
eqs.  (\ref{B.3})-(\ref{B.11}) should be substituted into
relations  (\ref{1.19}) and  (\ref{1.30a}).

Now we consider $A_m$-solutions with asymptotics (\ref{2.7}).
In this case all $w_1, \ldots, w_{m+1}$ are real and without
loss of generality $w_1 < \ldots < w_{m+1}$.
For $b_0^s= n_s$  from (\ref{2.20}) we get \cite{GrI}
\beq{B.12}
n_s = b_0^s = s(m-s+1),
\eeq
$s = 1, \ldots, m$, or explicitly
\beq{B.13}
b_0^1 = m, \quad b_0^2 = 2 (m -1), \ldots, b_0^m = m.
\eeq
>From (\ref{2.7}), (\ref{2.13}), $\bar{\mu} = \sqrt{C_1}$
and (\ref{B.3}) we get ($w_1 < \ldots < w_{m+1}$)
\bear{B.14}
\bar \mu b_0^1 = \bar \mu m = w_{m+1}, \\ \label{B.15}
\bar \mu b_0^2 = 2 \bar \mu(m-1) =  w_{m} + w_{m+1}, \\ \nn
\ldots \\  \label{B.16}
\bar \mu b_0^m = \bar \mu m =  w_{2}+ \dots + w_{m+1}.
\ear
These relations imply
\beq{B.17}
w_{m+1} = \bar \mu m, \quad  w_{m} = \bar \mu (m -2),
\ldots, w_{1} = - \bar \mu m,
\eeq
or,
\beq{B.18}
w_{j} = (2j - m -2) \bar \mu,
\eeq
$j = 1, \ldots, m + 1$. From (\ref{B.3}) and (\ref{B.17})
we get
\beq{B.19}
f_{s} = e^{-q^s} = \alpha_s^{(0)} e^{n_s \bar \mu u} +
\alpha_s^{(1)} e^{(n_s-2) \bar \mu u} +
\ldots +  \alpha_s^{(n_s)} e^{-n_s \bar \mu u},
\eeq
where $\alpha_s^{(k)}$ are constants, $k = 1, \ldots, n_s$,
$\alpha_s^{(n_s)} \neq 0$. Hence, due to (\ref{2.28}), (\ref{2.28a})
we obtain  the relations
\beq{B.19a}
H_{s} = e^{-q^s- n_s \bar \mu u } =
\alpha_s^{(0)}  + \alpha_s^{(1)} F + \ldots +  \alpha_s^{(n_s)} F^{n_s},
\eeq
equivalent to  (\ref{3.12})
($\alpha_s^{(0)}  + \alpha_s^{(1)} + \ldots +  \alpha_s^{(n_s)} =1)$
with $\alpha_s^{(n_s)} = P_s^{(n_s)}
\neq 0$, $s = 1, \ldots, m$. Thus, the Conjecture is proved for
the Lie algebras ${\cal G} = A_m$, $m \geq 1$.

Now we prove the Conjecture for
simple Lie algebras $C_{m+1}= sp(m+1)$, $m \geq 1$.
(Remind that for $m = 1$: $C_2 = B_2 = so(5)$).
The Cartan matrix for the Lie algebra
$C_{m+1}$ ($m \geq 1$) reads
\beq{C.1a}
\left(A_{ss'}\right)=
\left( \begin{array}{*{6}{c}}
2&-2&0&\ldots&0&0\\
-1&2&-1&\ldots&0&0\\
0&-1&2&\ldots&0&0\\
\multicolumn{6}{c}{\dotfill}\\
0&0&0&\ldots&2&-1\\
0&0&0&\ldots&-1&2
\end{array}
\right)\quad
\eeq
$s,s' = 0, \ldots, m$. The set of  equations  (\ref{3.1})
with the Cartan matrix (\ref{C.1a}) and
$s = 0, \ldots, m$, may be embedded into a set
of equations (\ref{B.1}) corresponding to the  Cartan matrix
of the Lie algebra $A_{2m+1}$ (see (\ref{B.1a}))
with  $s = -m, \ldots, 0, \ldots, m$, if the following
identifications : $\bar B_{-k} = \bar B_{k}$
and  $H_{-k} = H_{k}$, $k = 1, \ldots, m$, are adopted.
This proves the Conjecture for $C_{m+1}$,
since it was proved for $A_{2m+1}$.

\section{Some examples}

\subsection{Solution for $A_2$}

Here we consider some examples of solutions related
to the Lie algebra $A_2 = sl(3)$. According to the
results of previous section we  seek the solutions
to eqs. (\ref{3.1})-(\ref{3.2b}) in the following
form (see  (\ref{3.12}); here $n_1 = n_2 =2$):
\beq{4.1}
H_{s} = 1 + P_s z + P_s^{(2)} z^{2},
\eeq
where $P_s= P_s^{(1)}$ and $P_s^{(2)} \neq 0$ are constants,
$s = 1,2$.

The substitution of  (\ref{4.1}) into equations  (\ref{3.1})
and decomposition in powers of $z$ lead us to the relations
\bear{4.2}
 - P_s (P_s + 2 \mu )  + 2 P_s^{(2)} = \bar B_s, \\  \label{4.3}
 - 2 P_s^{(2)} (P_s + 4 \mu ) =  P_{s+1} \bar B_s, \\ \label{4.4}
 - 2 P_s^{(2)} (\mu P_s + P_{s}^{(2)}) =  P_{s+1}^{(2)} \bar B_s,
\ear
corresponding to powers $z^0, z^1, z^2$ respectively, $s = 1,2$.
Here we denote  $s+ 1 = 2, 1$ for $s = 1,2$ respectively.
For $P_1 +P_2 + 4\mu \neq 0$ the solutions of
(\ref{4.2})-(\ref{4.4}) read
\bear{4.5}
 P_s^{(2)} = \frac{ P_s P_{s +1} (P_s + 2 \mu )}{2 (P_1 +P_2 + 4\mu)},
 \\ \label{4.6}
\bar B_s = - \frac{ P_s (P_s + 2 \mu )(P_s + 4 \mu )}{P_1 +P_2 + 4\mu},
\ear
$s = 1,2$. For $P_1 +P_2 + 4\mu = 0$ there exist also a
special solution with
\bear{4.6a}
P_1= P_2 = -2 \mu, \qquad  2 P_{s}^{(2)} = \bar B_s >0,
\qquad \bar B_1 + \bar B_2 = 4 \mu^2.
\ear

Thus, in the $A_2$-case the solution is described by relations
(\ref{2.30})-(\ref{2.33}) with $S = \{s_1,s_2\}$,
intersection rules (\ref{1.40}), or, equivalently,
\bear{1.40a}
d(I_{s_1} \cap I_{s_2})= \frac{d(I_{s_1})d(I_{s_2})}{D-2}-
\chi_{s_1} \chi_{s_2} \lambda_{a_{s_1}}\cdot\lambda_{a_{s_2}}
- \frac12 K,
\\ \label{1.40b}
d(I_{s_i}) - \frac{(d(I_{s_i}))^2}{D-2}+
\lambda_{a_{s_i}}\cdot\lambda_{a_{s_i}} = K,
\ear
where
$K = K_{s_i} \neq 0$,  and functions $H_{s_i} = H_i$
are defined by relations
(\ref{4.1}) and (\ref{4.5})-(\ref{4.6a}) with $z = R^{-\bar d}$,
$i =1,2$.

\subsection{$A_2$-dyon in $D = 11$ supergravity}

Consider the  ``truncated''  bosonic sector of
$D=   11$ supergravity (``truncated''  means without
Chern-Simons term).  The action  (\ref{1.1}) in this
case reads  \cite{CJS}
\ber{4.7}
S_{tr} =   \int_{M} d^{11}z \sqrt{|g|}
\left\{ {R}[g] - \frac{1}{4!}  F^2 \right\}.
\eer
where ${\rm rank} F =   4$. In this particular case,
we consider a dyonic black-hole solutions
with  electric $2$-brane and magnetic  $5$-brane
defined on the manifold
\beq{4.8}
M =    (2\mu, +\infty )  \times
(M_1 = S^{2})  \times (M_2 = \R) \times M_{3} \times M_{4},
\eeq
where ${\dim } M_3 =  2$ and ${\dim } M_4 =  5$.

The solution reads,
\bear{4.9}
g=  H_1^{1/3} H_2^{2/3} \biggl\{ \frac{dR \otimes dR}{1 - 2\mu / R} +
R^2  d \Omega^2_{2} \\ \nn
 -  H_1^{-1} H_2^{-1} \left(1 - \frac{2\mu}{R} \right) dt\otimes dt
+ H_1^{-1} g^3 + H_2^{-1} g^4 \biggr\}, \mm
\label{4.10}
F =  - \frac{Q_1}{R^2} H_1^{-2} H_2  dR \wedge dt \wedge \tau_3+
Q_2 \tau_1 \wedge \tau_3,
\ear
where metrics $g^2$ and  $g^3$ are
Ricci-flat metrics of Euclidean signature,
and $H_s$  are defined as follows
\beq{4.11}
H_{s} = 1 + \frac{P_s}{R}+ \frac{P_s^{(2)}}{R^{2}},
\eeq
where parameters
$P_s$, $\mu > 0$ and $P_s^{(2)}$, $\bar B_s = B_s = - 2 Q_s^2$,
$s =1,2$, satisfy relations  (\ref{4.5}) and (\ref{4.6}).

The  solution describes $A_2$-dyon consisting
of electric  $2$-brane with world sheet isomorphic
to $(M_2 = \R) \times M_{3}$ and magnetic  $5$-brane
with worldsheet isomorphic to $(M_2 = \R) \times M_{4}$.
The ``branes'' are intersecting on the time manifold $M_2 = \R$.
Here  $K_s = (U^s,U^s)=2$, $\eps_s = -1$ for all $s \in S$.
The $A_2$ intersection rule reads (see (\ref{1.40}))
\beq{4.12}
2 \cap 5= 1
\eeq
Here and in what follows $(p_1 \cap p_2= d) \Leftrightarrow
(d(I)=p_1 + 1, d(J)= p_2 + 1, d(I\cap J) = d)$.

The solution (\ref{4.9}), (\ref{4.10})
satisfies not only equations of
motion for the truncated model,
but also  the equations of motion
for  $D =11$ supergravity with the bosonic sector action
\ber{4.13}
S =  S_{tr} +  c \int_{M} A \wedge F \wedge F
\eer
($c = {\rm const}$,  $F = d A$),
since the only modification
related to ``Maxwells'' equations
\ber{4.14}
d*F = {\rm const} \ F \wedge F,
\eer
is trivial due to $F \wedge F = 0$ (since $\tau_i \wedge \tau_i =0$).

This solution in a special case $H_1 = H_2 = H^2$
($P_1 = P_2$, $Q_1^2 =  Q_2^2$) was considered in \cite{CIM}.
The 4-dimensional section of the metric (\ref{4.9})
in this special case coincides
with the Reissner-Nordstr\"om  metric.
For the extremal case, $\mu \to + 0$,
and multi-black-hole generalization
see also  \cite{IMBl}.

\subsection{$A_2$-dyon in Kaluza-Klein model}

Let us  consider $4$-dimensional model
\beq{4.15}
S= \int_{M} d^4z \sqrt{|g|}\biggl\{R[g]- g^{\mu \nu}
\p_\mu \varphi \p_\nu \varphi
-\frac{1}{2!} \exp[2\lambda \varphi]F^2\biggr\}
\eeq
with scalar field  $\varphi$, two-form $F = d A$ and
\beq{4.16}
\lambda = - \sqrt{3/2}.
\eeq
This model originates after Kaluza-Klein reduction
of $5$-dimensional gravity. The 5-dimensional metric
in this case reads
\beq{4.16a}
g^{(5)} = \phi g_{\mu \nu} dx^{\mu} \otimes dx^{\nu}
		  + \phi^{-2} (dy + {\cal A}) \otimes (dy + {\cal A}),
\eeq
where
\beq{4.16b}
{\cal A} = \sqrt{2} A =  \sqrt{2} A_{\mu} dx^{\mu},
\qquad \phi = \exp(2 \varphi/\sqrt{6}).
\eeq

We consider the dyonic black-hole solution
carrying  electric  charge $Q_1$
and magnetic  charge $Q_2$, defined on the manifold
\beq{4.17}
M =    (2\mu, +\infty )  \times (M_1 =S^{2})  \times (M_2 = \R).
\eeq
This solution reads
\bear{4.18}
g= \left( H_1 H_2 \right)^{1/2}
\biggl\{ \frac{dR \otimes dR}{1 - 2\mu / R} +   R^2  d \Omega^2_{2}
 -  H_1^{-1} H_2^{-1} \left(1 - \frac{2\mu}{R } \right) dt\otimes dt
\biggr\}, \\ \label{4.19}
\exp(\varphi) = H_1^{\lambda/2} H_2^{- \lambda/2},
\\ \label{4.20}
F = dA = -  \frac{Q_1}{R^2} H_1^{-2} H_2  dR \wedge dt + Q_2 \tau_1,
\ear
where  functions $H_{s}$  are defined by relations (\ref{4.1}),
(\ref{4.5}) and (\ref{4.6})	with
$\bar B_s = - 2 Q_s^2$, $z = R^{- 1}$, $s =1, 2$;
where $\tau_1$ is volume form on $S^2$.

For 5-metric we obtain from (\ref{4.16a})-(\ref{4.19})
\bear{4.16c}
g^{(5)} = H_2 \biggl\{ \frac{dR \otimes dR}{1 - 2\mu / R}
+   R^2  d \Omega^2_{2}   -  H_1^{-1} H_2^{-1}
\left(1 - \frac{2\mu}{R } \right) dt\otimes dt \biggr\}
\\ \nn
 + H_1 H_2^{-1}  (dy + {\cal A}) \otimes (dy + {\cal A}),
\ear
$d {\cal A} = \sqrt{2} F$.

For $Q_2 \to 0$ we get the black hole version of
Dobiash-Maison solution from \cite{DoMa} and
for $Q_1 \to 0$ we are led to the black hole version of
Gross-Perry-Sorkin monopole solution from \cite{GrP,Sor},
see \cite{CGMS}.
The solution agrees with that of  \cite{Lee}.

\section{Post-Newtonian approximation}

Let $d_1 =   2$. Here we consider the 4-dimensional section of
the metric (\ref{2.30}), namely,
\bear{5.1}
g^{(4)} =   U \biggl\{ \frac{dR
\otimes dR}{1 - 2\mu / R} + R^2  d \Omega^2_{2} -
U_1 \left(1 - \frac{2\mu}{R} \right)  dt \otimes dt \biggr\},
\ear
in the range $R > 2\mu$, where
\bear{5.1a}
U =   \prod_{s \in S} H_s^{2 d(I_s) h_s/(D-2)},
\\  \label{5.1b}
U_1 =   \prod_{s \in S} H_s^{-2 h_s}.
\ear

Let us  imagine that some real astrophysical objects (e.g. stars) may
be described (or approximated)
by the 4-dimensional  physical  metric (\ref{5.1}),
i.e. they are   traces  of extended multidimensional objects
(charged $p$-brane black holes).

For post-Newtonian approximation
we restrict ourselves by the first two powers of $1/R$, i. e.
\ber{5.2a}
H_s = 1 + \frac{P_s}{R}  + \frac{P_s^{(2)}}{R^2} + o(\frac{1}{R^3}),
\eer
for $R \to + \infty$, $s \in S$.

Introducing a new radial variable $\rho$ by
the relation
\ber{5.2}
R =   \rho \left(1 + \frac{\mu}{2\rho}\right)^2,
\eer
($\rho > \mu/2$), we may rewrite the metric (\ref{5.1})
in the 3-dimensional conformally-flat form,
\ber{5.3}
g^{(4)} =   U \Biggl\{ -
U_1 \frac{\left(1 - \frac{\mu}{2 \rho} \right)^2}
{\left(1 + \frac{\mu}{2 \rho} \right)^2} dt \otimes dt +
\left(1 + \frac{\mu}{2 \rho} \right)^4
\delta_{ij} dx^i \otimes dx^j \Biggr\},
\eer
where $\rho^2 =  |x|^2 =   \delta_{ij}x^i x^j$ ($i,j =   1,2,3$).

For possible physical applications, one  should calculate the
post-Newtonian parameters $\beta$ and $\gamma$ (Eddington
parameters) using the following standard relations
\ber{5.4}
g^{(4)}_{00} =   - (1 -  2 V + 2 \beta V^2 ) + O(V^3),
\\
\label{5.5}
g^{(4)}_{ij} =   \delta_{ij}(1 + 2 \gamma V ) + O(V^2),
\eer
$i,j =   1,2,3$, where,
\ber{5.6}
V =   \frac{GM}{\rho}
\eer
is  Newton's potential, $G$ is the gravitational constant and
$M$ is the gravitational mass. From (\ref{5.1a})-(\ref{5.6}) we
deduce the formulas
 \ber{5.7}
GM =   \mu + \sum_{s \in S} h_s P_s
\left(1 -  \frac{d(I_s)}{D-2} \right)
\eer
and
\bear{5.8}
\beta - 1 =   \frac{1}{2(GM)^2} \sum_{s \in S}  h_s
(P_s^2 +2 \mu P_s - 2 P_s^{(2)})
\left(1 -  \frac{d(I_s)}{D-2} \right), \\
\label{5.9}
\gamma - 1 =   - \frac{1}{GM} \sum_{s \in S} h_s P_s \left(1 -  2
\frac{d(I_s)}{D-2} \right).
\ear

For special ``block-orthogonal'' solutions see \cite{CIM}.

Now, we show that in general case, like in \cite{CIM},
the parameter $\beta$ is defined by the charges squared
$Q_s^{2}$ of $p$-branes (or, more correctly, by the charge densities
squared) and signature parameters $\eps_s$. Indeed, from
(\ref{2.34}) we get in zeroth order of $1/R$-decomposition:
\bear{5.10}
P_s^2 +2 \mu P_s - 2 P_s^{(2)} = - B_s
\ear
with $B_s$ defined in (\ref{1.21}), $s \in S$.
Hence ,
\bear{5.11}
\beta - 1 =   \frac{1}{2(GM)^2} \sum_{s \in S} (- \eps_s) Q_s^{2}
\left(1 -  \frac{d(I_s)}{D-2} \right)
\ear

Here we succeeded in presenting of $\beta$
in terms of ratios of  physical parameters:
$Q_s/GM$. This parameter is obtained without knowledge
of the general solution for $H_s$ and does not depend
upon the quasi-Cartan matrix and, hence, upon intersections
of $p$-branes. The parameter $\gamma$
depends upon ratios $P_s/GM$, where $P_s$
are functions of $GM$, $Q_s$ and $A = (A_{s s'})$.
The calculation of $\gamma$ needs an exact solution
for radial functions $H_s$.

Relations (\ref{5.9}) and (\ref{5.11}) coincide
with those obtained in \cite{IMJ2,CIM}
for block-orthogonal case, when charges $Q_s$ (and $P_s$)
are  coinciding inside blocks.

For the most interesting from ``physical point' of view
$p$-brane solutions with $\eps_s = -1$ and  $d(I_s) < D - 2$
(for all $s \in S$) (\ref{5.11}) implies
\ber{5.12}
\beta > 1.
\eer

\section{Extremal case}

\subsection{``One-pole'' solution}

Here we consider the extremal case: $\mu \to +0$.
The relation  for the metric (\ref{2.30}) reads in this
case as follows
\bear{6.1}
g= \Bigl(\prod_{s \in S} H_s^{2 h_s d(I_s)/(D-2)} \Bigr)
\biggl\{ dR \otimes dR  + R^2  d \Omega^2_{d_1}  \\ \nn
-  \Bigl(\prod_{s \in S} H_s^{-2 h_s} \Bigr) dt \otimes dt
+ \sum_{i = 3}^{n} \Bigl(\prod_{s\in S}
  H_s^{-2 h_s \delta_{iI_s}} \Bigr) g^i  \biggr\},
\ear
and the relations for scalar fields and  fields
of forms (\ref{2.31})-(\ref{2.33}) are unchanged.
Here $H_s =  H_s(z) > 0$, $z = R^{-\bar d}$ and
the following relations  are satisfied
\bear{6.2}
 \frac{d^2}{dz^2} \ln H_s =
 \bar B_s \prod_{s' \in S}  H_{s'}^{- A_{s s'}},
 \qquad H_{s}(0) = 1, \\
\label{6.3}
E_{TL} = \frac{\bar{d}^2}{4}  \sum_{s,s' \in S} h_s
A_{s s'} (\frac{d}{dz}{\ln H_s}) \frac{d}{dz}{\ln H_{s'}}
+  \sum_{s \in S}   A_s \prod_{s' \in S}  H_{s'}^{- A_{s s'}} =0,
\ear
where $\bar B_s =  B_s/ \bar d^2 \neq 0$, and
$\bar B_s$, $A_s$ are defined in (\ref{1.21}), $s \in S$.
These solution may be  obtained as a special case of
solutions from Sect. 2 with
\beq{6.4}
C_1 =  E_{TL} = c^A = 0,
\eeq
$A = (i,\alpha)$ and $u = z/ \bar d$ (see (\ref{2.28a})).

{\bf Conjecture 2.} {\em Let $(A_{s s'})$ be  a Cartan matrix  for a
semisimple finite-dimensional Lie algebra. Then  the solution to
eqs. (\ref{6.2})-(\ref{6.3}) for $\bar B_s < 0$, $s \in S$, is
uniquely defined and is a polynomial
\beq{6.5}
H_{s}(z) = 1 + \sum_{k = 1}^{n_s} P_s^{(k)} z^k,
\eeq
where $P_s^{(k)}$ are constants,
$k = 1,\ldots, n_s$, integers $n_s = b_0^s$ are
defined in (\ref{3.11}) and $P_s^{(n_s)} \neq 0$, $s \in S$.}

This conjecture  was suggested (in fact) previously in \cite{LMMP}.
The coefficients $P_s^{(n_s)} = C_s > 0 $ may be calculated
by substitution of asymptotical relations
\beq{6.6}
 H_{s}(z) \sim  C_s z^{b_0^s},   \quad  z \to \infty
\eeq
into eqs.  (\ref{6.2}), (\ref{6.3}), $s \in S$. This results in
the relations
\beq{6.7}
C_s = \prod_{s' \in S} ( - b_0^{s'} \bar B_{s'})^{A^{s s'}},
\eeq
$s \in S$. We remind that $(A^{s s'}) = (A_{s s'})^{-1}$

We note that the asymptotical relations (\ref{6.6})
are satisfied in a more general case, when  $B_s b_0^s < 0$, $s \in S$.

Let us consider  the metric (\ref{6.1}) with $H_s$
obeying asymptotical relations (\ref{6.6}).
We have a horizon for $R  \to +0$, if
\bear{6.8}
\xi = \sum_{s\in S} h_s b^s_0 - \frac1{d_0-2}\ge0,
\ear
where $d_0 = d_1 +1$.
This relation follows from the requirement of infinite time
propagation of light to $R \to +0$.

For flat internal spaces  $M_i=\R^{d_i}$, $i=3,\dots,n$,
we get for the Riemann tensor squared (Kretschmann scalar) \cite{IMBl}
\beq{6.9}
R_{MNPQ}[g]R^{MNPQ}[g]= [C + o(1)] R^{4(d_0-2)\eta}
\eeq
for $R \to +0$, where
\ber{6.10}
\eta = \sum_{s \in S} h_s b^s_0 \frac{d(I_s)}{D-2}-
\frac{1}{d_0-2},
\eer
and $C \geq 0$ ($C = {\rm const}$).
Due to (\ref{6.9}) the metric (\ref{6.1})
with flat internal spaces has no
curvature singularity when $R \to + 0$, if
\bear{6.11}
\eta \ge 0.
\ear

For $h_s b^s_0 > 0$, $d(I_s)<D-2$, $s \in S $, we get
$\eta < \xi$ and relation (\ref{6.11}) single out extremal
charged  black hole  with $p$-branes and flat internal spaces.

\subsection{Multi-black-hole extension}

The solutions under consideration have a Majumdar-Papapetrou-type
extension defined on the manifold
\beq{6.12a}
M =    M_0  \times (M_2 = \R) \times  \ldots \times M_{n},
\eeq
The solution reads
\bear{6.13}
g= \Bigl(\prod_{s \in S} H_s^{2 h_s d(I_s)/(D-2)} \Bigr)
\biggl\{ g^0 -  \Bigl(\prod_{s \in S} H_s^{-2 h_s} \Bigr)  dt \otimes dt
+ \sum_{i = 3}^{n} \Bigl(\prod_{s\in S}
  H_s^{-2 h_s \delta_{iI_s}} \Bigr) g^i  \biggr\},
\\  \label{6.14}
\exp(\varphi^\alpha)=
\prod_{s\in S} H_s^{h_s \chi_s \lambda_{a_s}^\alpha},
\\  \label{6.15}
F^a= \sum_{s \in S} \delta^a_{a_s} {\cal F}^{s},
\ear
where
\beq{6.16}
{\cal F}^s= Q_s
\left( \prod_{s' \in S}  H_{s'}^{- A_{s s'}} \right) dH \wedge\tau(I_s),
\eeq
$s\in S_e$, and
\beq{6.17}
 {\cal F}^s= Q_s (*_0 d H) \wedge \tau(\hat I_s),
\eeq
$s\in S_m$. Here
\beq{6.17a}
\hat I \equiv \{2,\ldots,n\}\setminus I
\eeq
$g^0=g_{\mu\nu}^0(x)dx^\mu\otimes dx^\nu$
is a Ricci-flat metric on $M_0$ and $*_0 = *[g^0]$ is the
Hodge operator on $(M_0,g^0)$ and
\beq{6.18}
H_s =  H_s(H(x)),
\eeq
where functions $H_s =  H_s(z) > 0$, $z \in [0, +\infty)$, $s\in S$,
satisfy the relations (\ref{6.2}) and  (\ref{6.3}) and
$H = H(x) > 0$ is a harmonic function on $(M_0,g^0)$, i.e.
\beq{6.19}
\tri[g^0]H=0.
\eeq
Here $\tri[g^0]$ is the Laplace-Beltrami operator
corresponding to $g^0$. This solution
is a special case of the solutions from \cite{IK}
corresponding to restrictions (\ref{6.4}).

Let us consider as an example a flat space:
$M_0=\R^{d_0} \setminus X$, $d_0>2$ and
$g^0=\delta_{\mu\nu}dx^\mu\otimes
dx^\nu$  and
\beq{6.20}
H(x) = \sum_{b \in X}\frac{q_{b}}{|x- b|^{d_0-2}},
\eeq
where $X$ is finite non-empty subset $X \subset M_0$ and
all $q_{b}>0$ for $b \in X$. For flat internal spaces  $M_i=\R^{d_i}$,
$i=3,\dots,n$, and non-negative indices $\eta$ and $\xi$ (see (\ref{6.8})
and (\ref{6.11}))  the solution describes
a set of $|X|$ extremal $p$-brane black holes.
Here relations $H(x) \to 0$ for $|x| \to + \infty$
and $H_{s}(0) = 1$ ($s \in S$) imply
the asymptotical flatness of the $(1+d_0)$-dimensional section of the
metric. A black hole corresponding to a ``point'' (horizon) $b \in X$
carries brane charges $Q_s q_b$, $s \in S$.  Since the solution is
invariant under the replacement  of parameters:  $Q_s \mapsto \alpha Q_s $,
$q_b \mapsto  Q_s/\alpha $, $\alpha >0$, $b \in X$, $s \in S$,
we may normalize parameters $q_b$ by the restriction
\beq{6.21}
\sum_{b \in X} q_{b} = 1.
\eeq

\section{Conclusions}

Thus here we obtained a family
of black hole solutions  with intersecting $p$-branes
with next to arbitrary intersection rules,
see relations (\ref{1.17a}), (\ref{1.18b}) and
Restriction 1. (Restriction 2 is satisfied, since all
$p$-branes have a common time manifold.)
The metric of solutions  contains  $n -1$ Ricci-flat ``internal''
space metrics. The solutions are defined up to  a set of functions $H_s$
obeying a set of  equations (equivalent to Toda-type
equations) with certain boundary conditions imposed.
We suggested a conjecture on polynomial structure of  $H_s$
for intersections related to semisimple Lie algebras and
proved it for $A_m$ and $C_{m+1}$ algebras, $m \geq 1$..
We obtained explicit relations for the solutions in the $A_2$ case
and considered two examples of $A_2$-dyon solutions: one in $D =
11$ supergravity and another in $5$-dimensional
Kaluza-Klein theory. We also calculated
post-Newtonian parameters  $\beta$ and $\gamma$ corresponding to
$4$-dimensional section of the metric.
We  presented  $\beta$  as a function of ratios of  physical parameters:
charges $Q_s$ and mass $M$.
The parameter $\beta$ does not depend upon intersections
of $p$-branes. The parameter $\gamma$
depends upon intersections. Its calculation  needs an exact
solution for radial functions $H_s$. We also obtained
extremal black hole configurations.

\begin{center}
{\bf Acknowledgments}
\end{center}

This work was supported in part by the Russian Ministry for
Science and Technology, Russian Foundation for Basic Research,
and project SEE.


\small
\begin{thebibliography}{99}

\bibitem{M-th1}
E. Witten, {\it Nucl. Phys.} {\bf B 443}, 85 (1995); hep-th/9503124; \\
P. Townsend, {\it Phys. Lett. } {\bf B 350}, 184 (1995); hep-th/9612121; \\
C. Hull and P. Townsend, {\it Nucl. Phys.} {\bf B 438}, 109 (1995);
hep-th/9610167; \\
P. Horava  and E. Witten, {\it Nucl. Phys.} {\bf B 460}, 506 (1996);
hep-th/9510209.

\bibitem{M-th2}
J.M. Schwarz,  Lectures on Superstring and M-theory Dualities,
hep-th/9607201; \\
M.J. Duff,  M-theory (the Theory Formerly Known as Strings),
hep-th/9608117.


\bibitem{GSW}
M.B. Green, J.H. Schwarz and E. Witten, Superstring Theory,
vol. 1, 2, Cambridge, 1987.

\bibitem{St}
K.S. Stelle, Lectures on Supergravity p-branes, hep-th/9701088.

\bibitem{DKL}
M.J. Duff, R.R. Khuri and J.X. Lu,
{\it Phys. Rep.} {\bf 259},  213 (1995).

\bibitem{DGHR}
A. Dabholkar, G. Gibbons, J.A. Harvey, and F. Ruiz Ruiz,
{\it Nucl. Phys.} {\bf B 340}, 33 (1990).

\bibitem{HS}
G.T. Horowitz  and A. Strominger,
{\it Nucl. Phys.} {\bf B 360}, 197 (1990).

\bibitem{DS}
M.J. Duff and K.S. Stelle, {\it Phys. Lett.} {\bf B 253},  113 (1991).

\bibitem{Guv}
R. G\"{u}ven, Phys. Lett. {\bf B 276}, 49 (1992);
{\it Phys. Lett.} {\bf B 212}, 277 (1988).

\bibitem{Str}
A. Strominger, {\it Phys. Lett. } {\bf B 383}, 44 (1996);
hep-th/9512059.

\bibitem{To}
P.K. Townsend, {\it Phys. Lett. } {\bf B 373}, 68 (1996);
hep-th/9512062.

\bibitem{PT}
G. Papadopoulos and P.K. Townsend,
{\it  Phys. Lett.} {\bf B 380}, 273 (1996); hep-th/9603087.

\bibitem{Ts}
A.A. Tseytlin,
{\it Nucl. Phys.} {\bf B 475}, 149 (1996); hep-th/9604035.

\bibitem{GKT}
J.P. Gauntlett,  D.A. Kastor, and J. Traschen,
{\it Nucl. Phys.} {\bf B 478}, 544 (1996); hep-th/9604179.

\bibitem{CT}
M. Cvetic and A.A. Tseytlin,  Nucl. Phys. B 478, 181 (1996).

\bibitem{Ts1}
A.A. Tseytlin, {\it Nucl. Phys.} {\bf B 487}, 141 (1997);
hep-th/9609212.

\bibitem{LP}
H. L\"u, C.N. Pope, SL(N+1,R) Toda Solitons in Supergravities,
hep-th/9604058.

\bibitem{LPX}
H. L\"u, C.N. Pope, and K.W. Xu, Liouville and Toda Solitons in
M-theory, hep-th/9604058.

\bibitem{LMPX}
H. L\"u, S. Mukherji, C.N. Pope and K.-W. Xu,
Cosmological Solutions in String Theories,
hep-th/9610107.

\bibitem{V}
A. Volovich,  {\it  Nucl. Phys. }  {\bf B 487} (11), 141 (1997);
hep-th/9608095.

\bibitem{AV}
I.Ya. Aref'eva and A.I. Volovich,
{\it Class. Quantum Grav.} {\bf B 14}, 29901 (1997);
hep-th/9611026.

\bibitem{IM0}
V.D. Ivashchuk and V.N. Melnikov,
Intersecting p-brane Solutions in Multidimensional
Gravity and M-theory, hep-th/9612089;
{\it Grav. and Cosmol.} {\bf 2}, No 4, 204 (1996).

\bibitem{IM}
V.D. Ivashchuk and V.N. Melnikov,
{\it Phys. Lett. } {\bf B 403}, 23 (1997).

\bibitem{BREJS}
E. Bergshoeff, M. de Roo, E. Eyras, B. Janssen and
J.P. van der Schaar, {\it Class. Quantum Grav.} {\bf 14} , 2757 (1997);
hep-th/9612095.

\bibitem{AR}
I.Ya. Aref'eva and O.A. Rytchkov,
Incidence Matrix Description of Intersecting p-brane
Solutions, {\it Preprint} SMI-25-96, hep-th/9612236.

\bibitem{AEH}
R. Argurio, F. Englert and L. Hourant,
{\it Phys. Lett. } {B 398}, 2991 (1997); hep-th/9701042.

\bibitem{AIR}
I.Ya. Aref'eva, M.G. Ivanov and O.A. Rytchkov,
Properties of Intersecting p-branes in Various Dimensions,
{\it Preprint} SMI-05-97, hep-th/9702077.

\bibitem{AIV}
I.Ya. Aref'eva, M.G. Ivanov and I.V. Volovich,
Non-Extremal Intersecting p-Branes in Various Dimensions, hep-th/9702079;
{\it Phys. Lett. } {\bf B 406}, 44 (1997).

\bibitem{Oh}
N. Ohta, Intersection Rules for Non-extreme p-branes, hep-th/9702164.


\bibitem{IMC}
V.D. Ivashchuk and V.N. Melnikov,
Sigma-model for the Generalized  Composite p-branes,
hep-th/9705036; {\it Class. Quantum Grav.} {\bf 14}, 3001 (1997);
Corrigenda {\it ibid.} {\bf 15 } (12), 3941 (1998).

\bibitem{IMR}
V.D. Ivashchuk, V.N. Melnikov and M. Rainer,
Multidimensional $\sigma$-models with Composite Electric $p$-branes,
gr-qc/9705005; {\it Grav. and Cosmol. } {\bf 4}, No 1 (13), (1998).

\bibitem{BGIM}
K.A. Bronnikov, M.A. Grebeniuk, V.D. Ivashchuk and V.N. Melnikov,
Integrable Multidimensional Cosmology for
Intersecting $p$-branes,
{\it Grav. and Cosmol. } {\bf  3}, No 2(10), 105 (1997).

\bibitem{LMMP}
H. L\"u, J. Maharana, S. Mukherji  and C.N. Pope,
Cosmological Solutions, p-branes and the Wheeler De Witt
Equation,   hep-th/9707182.

\bibitem{GrIM}
M.A. Grebeniuk, V.D. Ivashchuk and V.N. Melnikov,
Integrable Multidimensional Quantum Cosmology  for Intersecting p-Branes,
{\it Grav. and Cosmol.\/} {\bf 3}, No 3 (11), 243 (1997), gr-qc/9708031.

\bibitem{BKR}
K.A. Bronnikov, U. Kasper and M. Rainer,
Intersecting Electric and Magnetic $p$-Branes: Spherically Symmetric
Solutions, gr-qc/9708058.

\bibitem{IMJ}
V.D. Ivashchuk  and  V.N. Melnikov, Multidimensional Classical
and Quantum Cosmology with Intersecting $p$-branes,
{\it J. Math. Phys.}, {\bf 39}, 2866 (1998); hep-th/9708157,

\bibitem{Y}
D. Youm, {\it Phys. Rept.}, {\bf 316} (1999) 1-232;  hep-th/9710046.

\bibitem{BIM}
K.A. Bronnikov, V.D. Ivashchuk and V.N. Melnikov,
The Reissner-Nordstr\"om Problem for
Intersecting Electric and Magnetic $p$-branes, gr-qc/9710054;
{\it Grav. and Cosmol.}, {\bf 3}, No 3(11), 203 (1997).

\bibitem{Br}
K.A. Bronnikov, Block-orthogonal Brane systems, Black
Holes and Wormholes, hep-th/9710207;
{\it Grav. and Cosmol.} {\bf 4}, No 1 (13),  49 (1998).

\bibitem{GR}
D.V. Gal'tsov and O.A. Rytchkov, Generating Branes via
Sigma models, hep-th/9801180.

\bibitem{IMBl}
V.D. Ivashchuk and V.N. Melnikov,
Madjumdar-Papapetrou Type Solutions in Sigma-model
and Intersecting p-branes,
{\it Class. Quantum Grav.} {\bf 16}, 849 (1999);
hep-th/9802121.

\bibitem{IKM}
V.D.Ivashchuk, S.-W.Kim and V.N.Melnikov, Hyperbolic Kac-Moody
Algebra from Intersecting $p$-branes, to appear in J. Math. Phys.;
hep-th/9803006.

\bibitem{GrI}
M.A. Grebeniuk and V.D. Ivashchuk,
Sigma-model Solutions and Intersecting p-branes
Related to Lie Algebras,
{\it Phys. Lett. } {\bf B 442}, 125 (1998);
hep-th/9805113.

\bibitem{Br2}
K.A. Bronnikov,
Gravitating Brane Systems: Some General Theorems,
gr-qc/9806102; to be published in {\it J. Math. Phys.}

\bibitem{GM1}
V.R. Gavrilov and V.N. Melnikov,
Toda Chains with Type $A_m$  Lie Algebra for Multidimensional Classical
Cosmology with Intersecting $p$-branes, In :  Proceedings of the
International seminar "Curent topics in mathematical cosmology", (Potsdam,
Germany , 30 March - 4 April 1998), Eds. M. Rainer and H.-J. Schmidt,
World Scientific, 1998,  p. 310; hep-th/9807004.

\bibitem{IMJ2}
V.D. Ivashchuk and V.N. Melnikov,
Multidimensional Cosmological and Spherically Symmetric Solutions
with Intersecting $p$-branes, gr-qc/9901001; \\
Cosmological and Spherically Symmetric Solutions
with Intersecting $p$-branes, {\it  J. Math. Phys.}
{\bf 40}, No 10 (1999).

\bibitem{CIM}
S. Cotsakis, V.D. Ivashchuk and V.N. Melnikov,
P-branes Black Holes and Post-Newtonian Approximation,
{\it Grav. and Cosmol.\/} {\bf 5}, No 1 (17), (1999); gr-qc/9902148.

\bibitem{GM2}
V.R. Gavrilov and V.N. Melnikov,
Toda Chains  Associated with   Lie Algebras  $A_m$ in Multidimensional
Gravitation and Cosmology with Intersecting $p$-branes,  submitted to
Theor. Math. Phys. (in Russian).

\bibitem{IK}
V.D. Ivashchuk and S.-W. Kim,
Solutions with intersecting p-branes related to Toda chains,
submit. to JMP; hep-th/9907019.

\bibitem{FS}
J. Fuchs and C. Schweigert, Symmetries, Lie algebras and
Representations. A graduate course for physicists
(Cambridge University Press, Cambridge, 1997).

\bibitem{CJS}
E. Cremmer, B. Julia, J. Scherk.
{\it Phys. Lett. } {\bf B 76}, 409 (1978).

\bibitem{MP}
S.D. Majumdar, {\it Phys. Rev. } {\bf 72}, 930 (1947);  \\
A. Papapetrou,  {\it Proc. R. Irish Acad. } {\bf A51}, 191 (1947).

\bibitem{IMZ}
V.D. Ivashchuk, V.N. Melnikov and A.I. Zhuk,
{\it Nuovo Cimento } {\bf B 104}, 575  (1989).

\bibitem{DoMa}
P. Dobiash and D. Maison,
{\it Gen. Rel. Grav.} {\bf 14}, 231 (1982).

\bibitem{GrP}
D.J. Gross and M.J. Perry,
{\it Nucl. Phys. } {\bf B 226}, 29  (1983).

\bibitem{Sor}
R.D. Sorkin, {\it Phys. Rev. Lett.} {\bf 51}, 87 (1983).

\bibitem{Lee}
S.-C. Lee, {\it Phys. Lett.} {\bf 149}, 98 (1984).

\bibitem{CGMS}
C.-M. Chen, D. V. Gal'tsov, K. Maeda and S. Sharakin,
{\it Phys. Lett.} {\bf B 453}, 7 (1999).

\bibitem{T}
M. Toda, {\it Progr. Theor. Phys.} {\bf 45}, 174 (1970).
Theory of Nonlinear Lattices (Springer-Verlag, Berlin, 1981).




\bibitem{And}
A. Anderson, {\it J. Math. Phys.} {\bf 37}, 1349 (1996);
hep-th/9507092.



\end{thebibliography}
\end{document}